\documentstyle[12pt]{article}

\begin{document}

\baselineskip 22pt

\begin{center}
{\Large \bf Ratios of $B$ and $D$ Meson Decay Constants\\
in Relativistic Quark Model}\\
\vspace{1.0cm}
Dae Sung Hwang$^1$ and Gwang-Hee Kim$^2$\\
{\it{Department of Physics, Sejong University, Seoul 133--747,
Korea}}\\
\vspace{2.0cm}
{\bf Abstract}\\
\end{center}

We calculate the ratios of $B$ and $D$ meson decay constants
by applying the variational method to the relativistic hamiltonian
of the heavy meson.
We adopt the Gaussian and hydrogen-type trial wave functions,
and use six different potentials of the potential model.
We obtain reliable results for the ratios, which are similar
for different trial wave functions and different potentials.
The obtained ratios show the deviation from the nonrelativistic
scaling law, and they are in a pretty good agreement with the
results of the Lattice calculations.
\\

\vfill

\noindent
PACS number(s): 12.39.Ki, 12.39.Pn, 13.20.Fc, 13.20.He,
13.25.Ft, 13.25.Hw,\\
\hspace*{3.25cm} 14.40.Nd\\

\noindent
$1$: e-mail: dshwang@phy.sejong.ac.kr\\
$2$: e-mail: gkim@cmp.sejong.ac.kr
\thispagestyle{empty}
\pagebreak

\baselineskip 22pt

The knowledge of the decay constant of the $B$ meson $f_B$ is
very important, since it affects the magnitude of $B-{\bar{B}}$
mixing and the size of $CP$ violation significantly.
There have been intensive theoretical and experimental
researches for improving its understanding.
However, its theoretical calculation is difficult because it is
in the realm of nonperturbative $QCD$ and the motion of
the light quark in $B$ meson is relativistic.
Understanding the decay constant better is also invaluable
because its information can reveal the inside structure of
the hadron.

Grinstein \cite{grin} observed that the double ratio of the decay
constants $(f_{B_s}/f_{B_d})$/\\
$(f_{D_s}/f_{D_d})$ is very close to
1 with small correction of order $m_s/m_Q$.
He calculated the double ratio with the heavy quark effective
theory, and obtained 0.967.
He remarked that the value of $f_{B_s}/f_{B_d}$,
which is an important factor for the relative strengths of
$B_s-{\bar{B}}_s$ and $B_d-{\bar{B}}_d$ mixings,
can then be given reliably from the knowledge of the measurable
$f_{D_s}/f_{D_d}$.
Oakes \cite{oakes} also calculated the double ratio based on the
assumption that chiral symmetry is broken by quark mass terms in
the Lagrangian.
He obtained its value as $1.004$, and emphasized the importance
of the fact that this double ratio is very close to 1.

When one treats the heavy-light meson in analogy with
the nonrelativistic situation, one expects the scaling law
$f_B / f_D \simeq {\sqrt{M_D / M_B}}$, since the reduced masses
of the light quark ($u$ or $d$ quark) in $B$ and $D$ mesons have
almost the same value, and $f_P^{\ 2}M_P=12\,
|\psi (0)|^2$ by the Van Royen-Weisskopf formula \cite{royen}
for the pseudoscalar meson $P$, where $\psi (0)$
is the wave function at origin of the relative motion of quarks
\cite{rosner}.
However, the light quark inside $B$ or $D$ meson has large
velocity, and its nonrelativistic treatment is not legitimate.
Indeed, our calculation of the decay constants
in the relativistic quark
model, which we present in this Letter, shows that the
nonrelativistic consideration is much deviated by
the relativistic motion of the light quark,
since this relativistic nature makes the $\psi (0)$ of $B$ and
$D$ mesons different appreciably.
This character of the relativistic motion has also been
exposed by the Lattice calculations
\cite{latt},
since they have obtained larger values of $f_B / f_D$
than ${\sqrt{M_D / M_B}}$ from the nonrelativistic scaling law.
This situation can be understood clearly through our relativistic
calculation.

The potential model has been successful for $\psi$ and $\Upsilon$
families with the nonrelativistic hamiltonian, since their heavy
quarks can be treated nonrelativistically.
However, for $D$ or $B$ meson it has been difficult to apply
the potential model because of the relativistic motion of the
light quark in $D$ or $B$ meson.
In our calculation we work with the purely relativistic hamiltonian,
and adopt the variational method \cite{hwang}.
We take the Gaussian and hydrogen-type wave functions separately
as trial wave functions \cite{greub}, and obtain the
ground state energy and wave function by minimizing the expectation
value of the relativistic hamiltonian.
By using the wave function at origin $\psi (0)$,
we can obtain the value of the decay constant $f_P$ from
the Van Royen-Weisskopf formula.
Then we can obtain the ratios of the decay constants.
The reason why we choose the Gaussian and hydrogen-type trial
wave functions is that the former one is appropriate to the
long range confining potential, and the latter one to the
short range asymptotically free potential.

The heavy-light pseudoscalar meson is composed of one heavy quark with
mass $m_Q$ and one light quark with $m_q$, and its relativistic
hamiltonian is
given by
\begin{equation}
H={\sqrt{{\bf p}^2+{m_Q}^2}}+{\sqrt{{\bf p}^2+{m_q}^2}}+V(r),
\label{b1}
\end{equation}
where ${\bf r}$ and ${\bf p}$ are the relative coordinate and its
conjugate momentum.
The hamiltonian in (\ref{b1}) represents the energy of the meson in
the center of mass coordinate, since in that reference frame the
momenta of both the heavy and light quarks have the same magnitude as
that of the conjugate momentum of the relative coordinate.

We apply the variational method to the hamiltonian (\ref{b1})
with the Gaussian and hydrogen-type trial wave functions.
The Gaussian wave function is given by
\begin{equation}
\psi ({\bf r})=({{\mu}\over {\sqrt{\pi}}})^{3/2}e^{-{\mu}^2r^2/2},
\label{b5}
\end{equation}
where $\mu$ is the variational parameter.
The Fourier transform of $\psi ({\bf r})$ gives the momentum space
wave function $\chi ({\bf p})$, which is also Gaussian,
\begin{equation}
\chi ({\bf p}) = {{1}\over {({\sqrt{\pi}}{\mu})^{3/2}}}
e^{-p^2/2{\mu}^2}.
\label{b6}
\end{equation}
The ground state is given by
minimizing the expectation value of $H$ in (\ref{b1}),
\begin{equation}
\langle H\rangle =\langle\psi\vert H\vert\psi\rangle =E(\mu ),
\ \ \
{{d}\over {d\mu }}E(\mu )=0\ \ {\rm{at}}\ \ \mu ={\bar{\mu}},
\label{b7}
\end{equation}
and then ${\bar{\mu}} \equiv p_{_F}$
represents the inverse size of the meson,
and $\bar E \equiv E({\bar{\mu}})$ its mass $M_P$ \cite{hwang}.
For the value of the light quark mass $m_q$ in (\ref{b1}), we use
the current quark mass given by
Dominguez and Rafael \cite{rafael}: $m_d=9.9$ MeV and $m_s=199$ MeV.

We perform the same calculation as the above for the hydrogen-type
wave function
\begin{equation}
\psi ({\bf r})={1\over {\sqrt{4\pi }}}{2\over a_0^{3/2}}
e^{-r/a_0},
\label{b5s}
\end{equation}
where $a_0$ is the variational parameter which
represents the size of the meson.
The momentum space wave function conjugate to the
$\psi ({\bf r})$ in (\ref{b5s}) is given by
\begin{equation}
\chi ({\bf p}) ={2{\sqrt{2}}\over \pi }
{a_0^{3/2}\over (a_0^2\, p^2+1)^2}.
\label{b6s}
\end{equation}
\vspace*{0.1cm}

For $V(r)$ in (\ref{b1}),
we consider the following six potentials of the potential model,
which we also display in Fig. 1.
We note in Fig. 1 the tendency that the potential which has higher
values of potential energy in the short range has lower values in the
long range, and vice versa.

\noindent
(A) Coulomb and linear potential of Eichten $et$ $al.$ \cite{eich}:
\begin{equation}
V(r)=-{{{\alpha}_c}\over {r}}+Kr,
\label{pot1}
\end{equation}
\hspace*{0.5cm}
${\rm{with}}\ \
{\alpha}_c=0.52,\ \ K=1/(2.34)^2\ {\rm{GeV}}^2,
\ \ m_c=1.84\ {\rm{GeV}},\ \ m_b=5.18\ {\rm{GeV}}.$

\vspace*{0.3cm}
\noindent
(B) Coulomb and linear potential of Hagiwara $et$ $al.$ \cite{hagi}:
\begin{equation}
V(r)=-{{{\alpha}_c}\over {r}}+Kr,
\label{pot2}
\end{equation}
\hspace*{0.5cm}
${\rm{with}}\ \
{\alpha}_c=0.47,\ \ K=0.19\ {\rm{GeV}}^2,\ \ m_c=1.32\ {\rm{GeV}},
\ \ m_b=4.75\ {\rm{GeV}}.$

\vspace*{0.3cm}
\noindent
(C) Power law potential of Martin \cite{martin}:
\begin{equation}
V(r)=-8.064\ {\rm{GeV}}\ +\ (6.898\ {\rm{GeV}})\ (r\cdot
1\ {\rm{GeV}})^{0.1},
\label{pot3}
\end{equation}
\hspace*{0.5cm}
${\rm{with}}\ \
m_c=1.8\ {\rm{GeV}},\ \ m_b=5.174\ {\rm{GeV}}.$

\vspace*{0.3cm}
\noindent
(D) Power law potential of Rosner $et$ $al.$ \cite{ros93}:
\begin{equation}
V(r)=-0.772\ {\rm{GeV}}\ +\
0.801 \ (\ (r\cdot 1\ {\rm{GeV}})^{\alpha}\ -\ 1\ )\ /\ \alpha ,
\label{pot4}
\end{equation}
\hspace*{0.5cm}
${\rm{with}}\ \
\alpha = -0.12,\ \
m_c=1.56\ {\rm{GeV}},\ \ m_b=4.96\ {\rm{GeV}}.$

\vspace*{0.3cm}
\noindent
(E) Logarithmic potential of Quigg and Rosner \cite{qr}:
\begin{equation}
V(r)=-0.6635\ {\rm{GeV}}\ +\ (0.733\ {\rm{GeV}})\ {\rm{log}}(r\cdot
1\ {\rm{GeV}}),
\label{pot5}
\end{equation}
\hspace*{0.5cm}
${\rm{with}}\ \
m_c=1.5\ {\rm{GeV}},\ \ m_b=4.906\ {\rm{GeV}}.$

\vspace*{0.3cm}
\noindent
(F) Richardson potential \cite{richard}:
\begin{equation}
V(r)={8\pi \over 33-2n_f}\Lambda
(\Lambda r - {f(\Lambda r) \over \Lambda r}),\
f(t) = 1 - 4\int_1^{\infty}{dq\over q}
{{\rm{e}}^{-qt} \over [{\rm{ln}}(q^2-1)]^2+{\pi}^2},
\label{pot6}
\end{equation}
\hspace*{0.5cm}
${\rm{with}}\ \
n_f=3,\ \ \Lambda =0.398\ {\rm{GeV}},\ \
m_c=1.491\ {\rm{GeV}},\ \ m_b=4.884\ {\rm{GeV}}.$
\vspace*{0.3cm}

The results of the variational calculations with the Gaussian wave
function are organized in Table 1, and those with the hydrogen-type
wave function in Table 2.
We see in Table 1 and 2 that the larger energy ($\bar{E}$) state
has the smaller size of meson ($1/{\bar{\mu}}$ or $a_0$).
In order to check whether the Gaussian wave function in (\ref{b5})
is a really good wave function,
we enlarged the trial wave function by adding the second
excited harmonic oscillator eigenfunction which is an even function,
since the first excited one which is an odd function can not be
included in the ground state wave function of the relativistic
hamiltonian which commutes with the parity operator.
For this enlarged trial wave function we obtained the result that
the Gaussian part contributes much dominantly, therefore it confirms
that the Gaussian
wave function is a very good trial wave function in the variational
calculation of the relativistic hamiltonian in (\ref{b1}) with the
harmonic type wave function.
We also checked the wave function in (\ref{b5s}) by adding the first
excited hydrogen-type wave function, and confirmed that the wave
function in (\ref{b5s}) is also much dominant for the ground state
wave function of (\ref{b1}).

The decay constant $f_P$ of the pseudoscalar meson $P$ is defined by
the matrix element $\langle 0 | A_{\mu} | P(q)\rangle$:
\begin{equation}
\langle 0 | A_{\mu} | P(q)\rangle = i q_{\mu} f_P .
\label{b10}
\end{equation}
By considering the low energy limit of the heavy meson annihilation,
we have the relation between $f_P$ and the ground state wave function
at origin ${\psi}_P(0)$ from the Van Royen-Weisskopf formula
with the color factor \cite{royen}:
\begin{equation}
f_P^2={12\over M_P}|{\psi}_P(0)|^2,
\label{b11}
\end{equation}
where $M_P$ is the heavy meson mass.
Using this formula, from (\ref{b5}) and (\ref{b5s}) we have
\begin{eqnarray}
f_P&=&\sqrt{{12\over M_P}} \,\,\,
{\Bigl( }\, {p_{_F}(P)\over {\sqrt{\pi}}}\, {\Bigr) }^{3/2}
\ \ \ \ \ \ \ \
{\rm for\ Gaussian\ wave\ function},
\label{b12a}\\
   &=&\sqrt{{12\over M_P}} \,\,\,
{\Bigl( }{1\over {\pi}^{1/3} a_0(P)}{\Bigr) }^{3/2}
\ \ \ \
{\rm for\ hydrogen-type\ wave\ function}.
\label{b12b}
\end{eqnarray}
Then the ratio of the decay constants of pseudoscalar mesons $A$
and $B$ is given by \cite{hwang}
\begin{eqnarray}
{f_A\over f_B}&=&
\sqrt{{M_B\over M_A}}\times
\Bigl( \, {p_{_F}(A)\over p_{_F}(B)}\, \Bigr)^{3/2}\ \ \ \ \
{\rm for\ Gaussian\ wave\ function},
\label{b13a}\\
              &=&
\sqrt{{M_B\over M_A}}\times
\Bigl( \, {a_0(B)\over a_0(A)}\, \Bigr)^{3/2}\ \ \ \ \
{\rm for\ hydrogen-type\ wave\ function}.
\label{b13b}
\end{eqnarray}

By using the values in Table 1 and 2 for $M$ ($\bar{E}$),
$p_{_F}$ ($\bar{\mu}$), and $a_0$
in (\ref{b13a}) and (\ref{b13b}),
we obtain the ratios of the decay constants
for the potential models (A)-(F), which we present in Table 3 and 4.
We see in Table 3 and 4
that $f_{B_s} / f_{D_s}$ is enhanced, compared with the
nonrelativistic scaling law $\sqrt{M_{D_s} / M_{B_s}}$ whose
experimental value is 0.605,
by the factor of 1.324
which is induced by the factor of $|\psi_{B_s}(0)/\psi_{D_s}(0)|$.
For the estimation of this enhancement factor we used the average
of the fourth columns of Table 3 and 4.
Sometimes this factor has been approximated to be 1 and
$f_{B_s}/f_{D_s}\simeq \sqrt{M_{D_s} / M_{B_s}}$ has been used,
by treating it in analogy with the nonrelativistic case \cite{rosner}.
However our calculation shows that this factor is indeed important
and different from 1 significantly.
The same situation happens for the ratio $f_{B_d} / f_{D_d}$.
Compared with $\sqrt{M_{D_d} / M_{B_d}}$ whose experimental value is
0.595,
$f_{B_d} / f_{D_d}$ is enhanced by the factor of 1.335,
which is estimated from the average of the fifth columns of Table 3
and 4.
This situation is in agreement with the results of the Lattice
calculations \cite{latt}, as we see in Table 5,
where we organized the results of Lattice calculations.

As an application of the obtained ratios of decay constants,
let us consider the determination of the values of
$f_{D_d}$, $f_{B_s}$,
and $f_{B_d}$ from the experimental value of $f_{D_s}$.
The cleanest way to obtain the value of the decay constant from the
experimental results is through the purely leptonic decays of the
$D_s^+$ meson,
which are understood theoretically to occur via an annihilation of the
two valence quarks.
The decay rate of the $D_s^+$ meson is given by the formula
\cite{rosbanff}
\begin{equation}
\Gamma (D_s^+\rightarrow l^+\nu )\, =\, {1\over 8\pi}\, G_F^2\,
f_{D_s}^2\, m_l^2\, M_{D_s}\,
{\Bigl( }1-{m_l^2\over M_{D_s}^2}{\Bigr) }^2\, |V_{cs}|^2,
\label{b16}
\end{equation}
where $f_{D_s}$ is the meson decay constant, $M_{D_s}$ is the $D_s$
mass, $m_l$ is the mass of the final-state lepton,
$G_F$ is the Fermi coupling constant, and $V_{cs}$ is the CKM matrix
element.
The WA75 and CLEO collaborations took the data for the branching ratio
$B(D_s^+\rightarrow {\mu}^+\nu )$ \cite{wa75,cleo94}, and the Review
of Particle Properties \cite{rpp} presents
$B(D_s^+\rightarrow {\mu}^+\nu )=(5.9\pm 2.2)\times 10^{-3}$.
By using this branching ratio, the life time of $D_s$ meson
$\tau =(0.467\pm 0.017)\times 10^{-12}$ s,
$M_{D_s}=1968.5\pm 0.7$ MeV,
$m_{\mu}=105.66$ MeV,
$G_F=1.1664\times 10^{-5}$ ${\rm{GeV}}^{-2}$,
and $|V_{cs}|=1.01\pm 0.18$ \cite{rpp},
we obtain the following value of $f_{D_s}$ from (\ref{b16}):
\begin{equation}
f_{D_s}\ =\ 265\pm 68\ {\rm{MeV}}.
\label{b17}
\end{equation}
The uncertainty of the value of $f_{D_s}$ is due to those of the
experimental values of $B(D_s^+\rightarrow l^+\nu )$ and $|V_{cs}|$,
therefore if their values are improved experimentally, we can obtain
$f_{D_s}$ very accurately.
When we combine the $f_{D_s}$ value in (\ref{b17})
and the ratios given by the average of those in
Table 3 and 4, we get the following values of $f_{D_d}$, $f_{B_s}$,
and $f_{B_d}$:
$f_{D_d}=253\pm 65\ {\rm{MeV}}$,
$f_{B_s}=212\pm 54\ {\rm{MeV}}$,
$f_{B_d}=201\pm 51\ {\rm{MeV}}$.

In summary, we calculated the various ratios of the $B$ and $D$
meson decay constants by applying the variational method to the
relativistic hamiltonian.
We took the Gaussian and hydrogen-type trial wave functions
separately, and used six different potentials for the potential
energy.
We obtained the results which are similar for different trial
wave functions and different potentials.
This fact implies that our method is reliable.
The obtained results for the ratio $f_B/f_D$ show that the
nonrelativistic scaling law $f_B/f_D \simeq \sqrt{M_D/M_B}$
should be implemented by the relativistic consideration.
Its enhancement factor we obtained is about 1.33, which is
induced by $|\psi_{B}(0)/\psi_{D}(0)|$.
This result is in a pretty good agreement with the recent
Lattice calculations.
Our results for $f_{B_s}/f_{B_d}$ and $f_{D_s}/f_{D_d}$
are both about 1.05.
We also obtained the value of the double ratio
$(f_{B_s}/f_{B_d}) / (f_{D_s}/f_{D_d})$
as 1.008,
whereas Grinstein obtained 0.967 with the heavy quark effective
theory, and Oakes 1.004 with the chiral symmetry breaking.
\\

\noindent
{\em Acknowledgements} \\
\indent
This work was supported
in part by the Basic Science Research Institute Program,
Ministry of Education, Project No. BSRI-94-2414,
and in part by Daeyang Foundation at Sejong University in 1994.\\

\pagebreak

\pagebreak

\begin{table}[h]
\begin{center}
\begin{tabular}{|c|c|c|c|c|c|c|c|c|}   \hline
Model&${\bar{\mu}}(B_s)$&${\bar{E}}(B_s)$
&${\bar{\mu}}(B_d)$&${\bar{E}}(B_d)$
&${\bar{\mu}}(D_s)$&${\bar{E}}(D_s)$
&${\bar{\mu}}(D_d)$&${\bar{E}}(D_d)$\\   \hline
A ( Eich. )&0.565&5.933&0.544&5.896&0.495&2.661&0.478&2.620\\
B ( Hagi. )&0.550&5.553&0.530&5.516&0.466&2.217&0.451&2.174\\
C (Power 1)&0.587&5.154&0.567&5.119&0.510&1.855&0.493&1.815\\
D (Power 2)&0.600&5.353&0.577&5.318&0.495&2.042&0.475&2.000\\
E ( Log.  )&0.586&5.396&0.565&5.360&0.490&2.080&0.472&2.038\\
F ( Rich. )&0.584&5.404&0.564&5.368&0.495&2.103&0.479&2.062\\
\hline
(Average)  &0.579&5.466&0.558&5.430&0.492&2.160&0.475&2.118\\
\hline
\end{tabular}
\end{center}
\caption{The values of the variational parameter $\mu$ which
minimize $\langle H\rangle$,
and the corresponding values of the minimum
energy for the Gaussian wave function}
\end{table}

\begin{table}[h]
\begin{center}
\begin{tabular}{|c|c|c|c|c|c|c|c|c|}   \hline
Model&$a_0(B_s)$&${\bar{E}}(B_s)$
&$a_0(B_d)$&${\bar{E}}(B_d)$
&$a_0(D_s)$&${\bar{E}}(D_s)$
&$a_0(D_d)$&${\bar{E}}(D_d)$\\   \hline
A ( Eich. )&1.260&5.884&1.310&5.845&1.549&2.140&1.605&2.094\\
B ( Hagi. )&1.309&5.512&1.360&5.472&1.570&2.189&1.625&2.143\\
C (Power 1)&1.305&5.118&1.353&5.078&1.504&1.823&1.561&1.779\\
D (Power 2)&1.264&5.303&1.317&5.264&1.532&1.997&1.602&1.951\\
E ( Log.  )&1.308&5.200&1.360&5.160&1.593&1.876&1.660&1.829\\
F ( Rich. )&1.260&5.359&1.307&5.321&1.503&2.068&1.558&2.024\\
\hline
(Average)  &1.284&5.396&1.335&5.357&1.542&2.016&1.602&1.970\\
\hline
\end{tabular}
\end{center}
\caption{The values of the variational parameter $a_0$ which
minimize $\langle H\rangle$,
and the corresponding values of the minimum
energy for the hydrogen-type wave function.}
\end{table}

\pagebreak

\begin{table}[h]
\begin{center}
\begin{tabular}{|c|c|c|c|c|c|}   \hline
Model&$f_{B_s}/f_{B_d}$&$f_{D_s}/f_{D_d}$&
${(f_{B_s}/f_{B_d})\over (f_{D_s}/f_{D_d})}$&
$f_{D_s}/f_{B_s}$&$f_{D_d}/f_{B_d}$\\  \hline
A ( Eich. )&1.055&1.046&1.009&1.225&1.236\\
B ( Hagi. )&1.054&1.040&1.013&1.234&1.250\\
C (Power 1)&1.050&1.041&1.009&1.350&1.362\\
D (Power 2)&1.057&1.053&1.004&1.213&1.218\\
E ( Log.  )&1.053&1.047&1.005&1.232&1.238\\
F ( Rich. )&1.050&1.040&1.010&1.251&1.263\\   \hline
 (Average) &1.053$\pm .003$&1.045$\pm .005$&1.008$\pm .003$
&1.251$\pm .046$&1.261$\pm .047$\\
              \hline
\end{tabular}
\end{center}
\caption{Ratios of the decay constants obtained for the
Gaussian wave function.}
\end{table}

\begin{table}[h]
\begin{center}
\begin{tabular}{|c|c|c|c|c|c|}   \hline
Model&$f_{B_s}/f_{B_d}$&$f_{D_s}/f_{D_d}$&
${(f_{B_s}/f_{B_d})\over (f_{D_s}/f_{D_d})}$&
$f_{D_s}/f_{B_s}$&$f_{D_d}/f_{B_d}$\\  \hline
A ( Eich. )&1.057&1.043&1.013&1.217&1.232\\
B ( Hagi. )&1.055&1.042&1.013&1.208&1.223\\
C (Power 1)&1.052&1.044&1.007&1.354&1.363\\
D (Power 2)&1.060&1.057&1.003&1.221&1.224\\
E ( Log.  )&1.056&1.050&1.006&1.239&1.246\\
F ( Rich. )&1.053&1.044&1.008&1.236&1.246\\   \hline
 (Average) &1.056$\pm .003$&1.047$\pm .005$&1.008$\pm .004$
&1.246$\pm .050$&1.256$\pm .049$\\
              \hline
\end{tabular}
\end{center}
\caption{Ratios of the decay constants obtained for the
hydrogen-type wave function.}
\end{table}

\begin{table}[h]
\vspace*{0.7cm}
\hspace*{-1.7cm}
\begin{tabular}{|c|c|c|c|c|c|}   \hline
Group&$f_{B_s}/f_{B_d}$&$f_{D_s}/f_{D_d}$&
${(f_{B_s}/f_{B_d})\over (f_{D_s}/f_{D_d})}$&
$f_{D_s}/f_{B_s}$&$f_{D_d}/f_{B_d}$\\  \hline
 ELC  \cite{ELC} &1.08$\pm .06$&1.08$\pm .02$&1.00$\pm .06$
&1.03$\pm .22$&1.02$\pm .21$\\

UKQCD\cite{UKQCD}&1.22${\, }^{+.04}_{-.03}$&1.18$\pm .02$
&1.03${\, }^{+.04}_{-.03}$&1.09${\, }^{+.04+.42}_{-.03-.06}$
&1.16${\, }^{+.05+.46}_{-.05-.14}$\\

 BLS  \cite{BLS} &1.11$\pm .02\pm .05$&1.11$\pm .02\pm .05$
&1.00$\pm .03\pm .06$&1.11$\pm .06\pm .27$&1.11$\pm .08\pm .30$\\

 MILC \cite{MILC}&1.13(2)(9)(4)&1.09(1)(4)(4)&1.04(2)(9)(5)
&1.18(3)(17)(13)&1.22(5)(17)(19)\\
                    \hline
\end{tabular}
\caption{The results of the ratios from Lattice calculations:
the third column was estimated from the values of the first
and second columns, and the fourth and fifth columns were
estimated from the Lattice calculation results of
$f_{D_s}$, $f_{B_s}$, $f_{D_d}$, and $f_{B_d}$.}
\end{table}

\pagebreak

\vspace*{3.5cm}
\vspace*{14.5cm}
\noindent
Fig. 1. The potentials (A)-(F)
given in Eqs. (\ref{pot1})-(\ref{pot6}).
The radial distance of the horizontal axis is in the unit of
${\rm{GeV}}^{-1}$ (1 ${\rm{GeV}}^{-1}$ = 0.197 fm),
and the potential energy of the vertical axis is in the unit
of GeV.

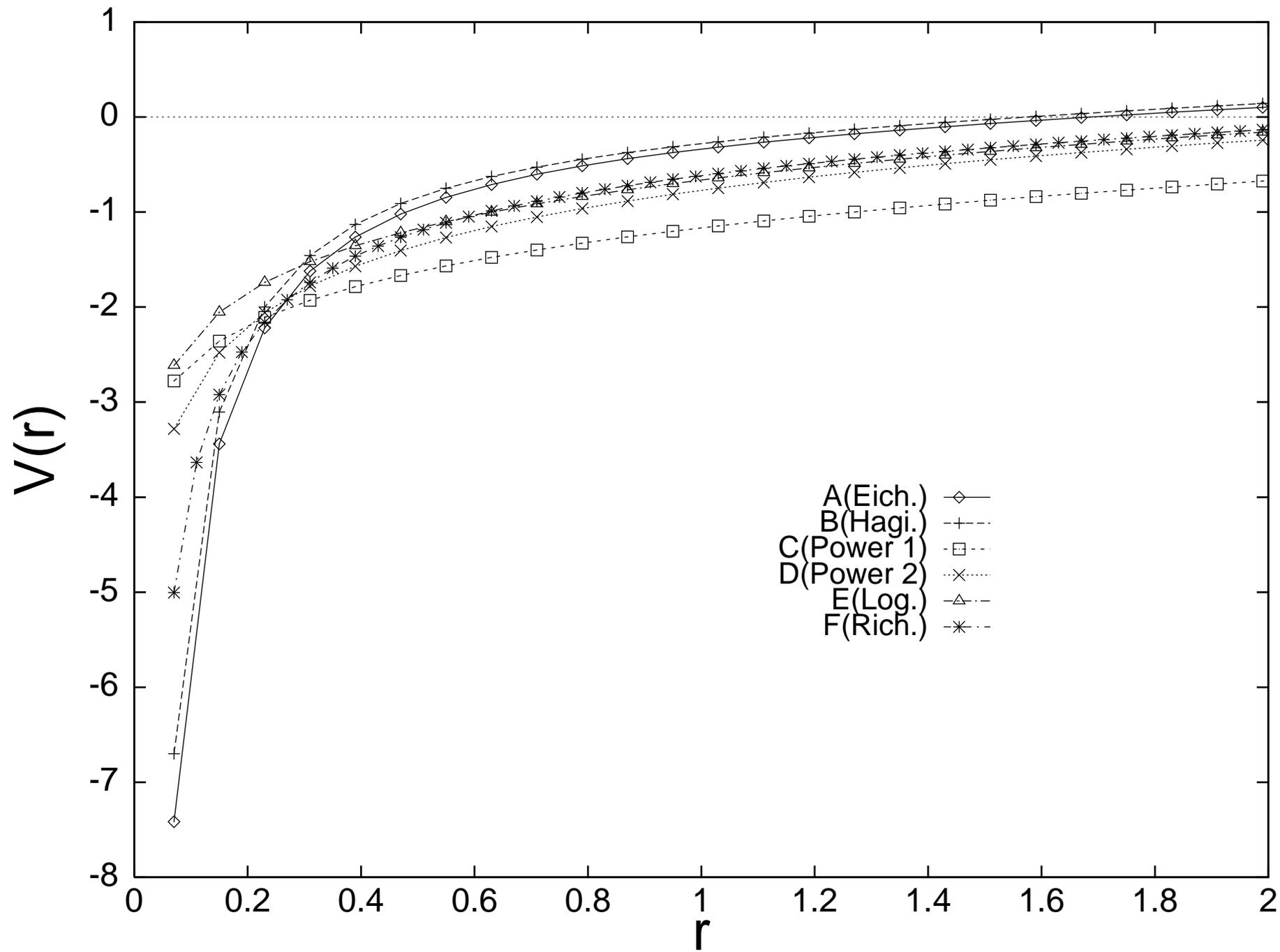
\end{document}